\documentclass[]{spie}  

\usepackage{amsmath,amsfonts,amssymb}
\usepackage{graphicx}
\usepackage[colorlinks=true, allcolors=blue]{hyperref}

\title{NIRPS Front-End: Design, performance, and lessons learned}
\authorinfo{Further author information, send correspondence to N. Blind (nicolas.blind@unige.ch)}

\author[a]{N. Blind}
\author[a,b]{U. Conod}
\author[c]{A. de Meideros}
\author[a]{F. Wildi}
\author[a]{F. Bouchy}
\author[a]{S. Bovay}
\author[d]{D. Brousseau}
\author[e]{A. Cabral}
\author[a]{L. Genolet}
\author[f]{J. Kolb}
\author[a]{R. Schnell}
\author[a]{A. Segovia}
\author[a]{M. Sordet}
\author[d]{S. Thibault}
\author[g]{B. Wehbe}
\author[f]{G. Zins}

\affil[a]{Observatoire astronomique de l'Universit\'e de Gen\`eve, 51 Chemin de Pegasi, CH-1290 Sauverny, Switzerland}
\affil[b]{Department of Physics and Astronomy, University of British Columbia, 6224 Agricultural Rd., Vancouver, BC, Canada}
\affil[c]{Departamento de Engenharia El\'etrica, Universidade Federal do Rio Grande do Norte, 59072-970 Natal, RN, Brazil}
\affil[d]{D\'epartement de physique, de g\'enie physique et d’optique, Universit\'e Laval, Qu\'ebec, QC, Canada G1V 0A6}
\affil[e]{Instituto de Astrofísica, Universidade de Lisboa, Estrada do Pa\c{c}o do Lumiar 22, PT1649-038 Lisboa, Portugal}
\affil[f]{European Southern Observatory, Karl-Schwarzschild-Straße 2, 85748 Garching bei München, Germany}
\affil[g]{Instituto de Astrof\'isica e Ci\^encias do Espa\c{c}o, Universidade do Porto, CAUP, Rua das Estrelas, 4150-762 Porto, Portugal}

\begin{document} 
\maketitle

\begin{abstract}
NIRPS (Near Infra-Red Planet Searcher) is an AO-assisted and fiber-fed spectrograph for high precision radial velocity measurements in the YJH-bands. NIRPS also has the specificity to be an SCAO assisted instrument, enabling the use of few-mode fibers for the first time. This choice offers an excellent trade-off by allowing to design a compact cryogenic spectrograph, while maintaining a high coupling efficiency under bad seeing conditions and for faint stars. The main drawback resides in a much more important modal-noise, a problem that has to be tackled for allowing 1m/s precision radial velocity measurements.
In this paper, we present the NIRPS Front-End: an overview of its design (opto-mechanics, control), its performance on-sky, as well as a few lessons learned along the way.
\end{abstract}


\section{Introduction}
\label{sec:intro} 
NIRPS (Near Infra-Red Planet Searcher) is a precision Radial Velocity (pRV) instrument mounted at the 3.6m-ESO telescope at La Silla Observatory. It will be operated together with HARPS and will cover the near Infra-Red from 980nm to 1800nm.  NIRPS has the specificity to be an SCAO-assisted instrument, allowing the main fiber (so called High Accuracy fibers (HAF)) to be only 0.4" in diameter, and the spectrograph and grating to be much more compact than seeing-limited instruments. A seeing limited fiber (High Efficiency Fiber (HEF)) has 0.9" is nevertheless available delivering 20\% lower spectral resolution.

The Front-End and AO system was installed and commissioned in December 2019. 4 commissioning runs happened since then. The Back-End was delivered to La Silla in February 2022, before First light of the complete instrument in April 2022. NIRPS is the only planned NIR pRV spectrograph to be installed in the Southern hemisphere.

NIRPS science case focuses on finding and confirming earth-mass planets in the habitable zone of low-mass M stars, in particular those identified by future space missions like TESS and PLATO. Such small planets require a radial velocity follow-up at a precision better than 1m/s. The simultaneous observation with HARPS, covering the 400-1800nm domain, will in addition help disentangle the stellar activity signal from planetary RV signal. This will be particularly important for early-to-mid-M dwarfs, where NIRPS and HARPS are expected to have a similar RV accuracy. NIRPS will also be used for the atmosphere characterisation of known transiting exoplanets, through very high-resolution transit spectroscopy.

In this paper, we present the NIRPS Front-End, giving an overview of its design (opto-mechanics, control), presenting its performance on-sky, and concluding on a few lessons learned along the way.

\section{Front-End Design}
\subsection{Overview}
\begin{figure}
    \centering
    \includegraphics[width=0.7\textwidth]{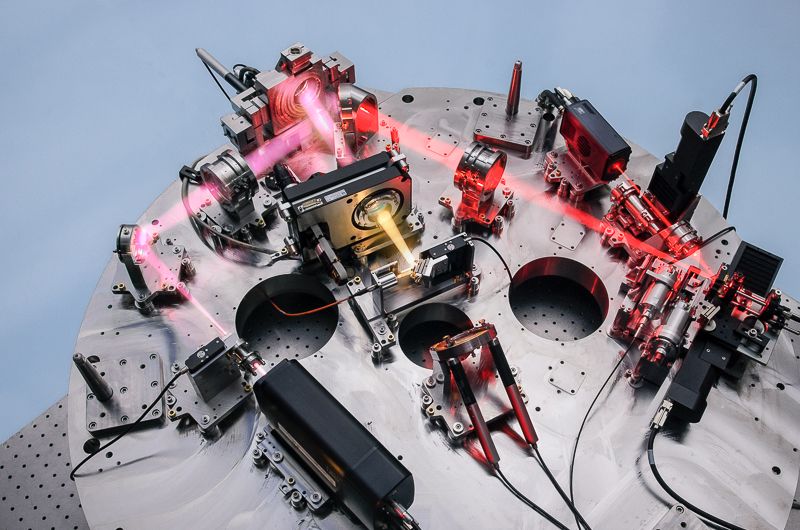}
    \caption{View of the NFE (NIRPS part only) at the end of integrations in Geneva in 2019. Optical path is visible from FP1 calibrations sources up to the different focal points.}
    \label{fig:FE_pic}
\end{figure}

\begin{figure}
    \centering
    \includegraphics[width=17cm]{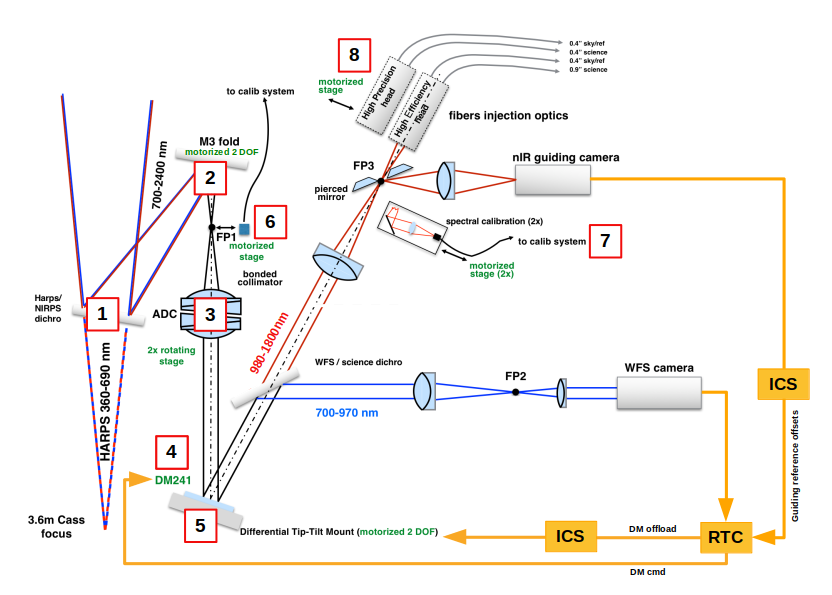}
    \caption{Schematics of NIRPS Front-End showing the optical layout, opto-mechanical functions as well as AO and guiding loops.}
    \label{fig:FE_scheme}
\end{figure}

Fig~\ref{fig:FE_pic} shows the NIRPS Front-End (NFE) top-bench at the end of its integration, while Fig.~\ref{fig:FE_scheme} shows a schematics of its opto-mechanical functions and control loops. The NFE is first responsible to split light between HARPS and NIRPS. The first dichroic [1] transmits light (380-690nm) to HARPS. It was verified during commissioning that this dichroic is nearly transparent to HARPS, introducing no RV offset in particular. It is removed in case of HARPS-POL observations.
The reflected, NIR part ensures functionalities common to many SCAO systems, distributing light between the WFS (700-950nm) and the science fibers (980-1800nm). We did not implement a cold-stop in the Front-End, thermal emission being negligible up to 1800nm, even with the HEF (0.9") fiber head. 

The role of the different motorized stages are the following:
\begin{enumerate}
    \item Moves the VIS/NIR dichroic in/out to enable HARPS-only observations, especially in case of HARPS-POL mode.
    \item M3 fold mirror is a Pupil Tip-Tilt Mirror (PTTM) used to center the telescope pupil on the WFS (see Sect.~\ref{sec:adc});
    \item Atmospheric dispersion corrector (ADC) covering the full bandwidth 700-1800 nm (see Sect.~\ref{sec:adc});
    \item ALPAO DM241 deformable mirror;
    \item Tip-tilt mount (TTM) that supports the DM and allows fine acquisition after telescope preset, with stroke $\ge \pm 40$'';
    \item Moveable FP1 point source connected to 940nm and 1550nm lasers for WFS and NCPA calibrations.
    \item Calibration light: 2 identical opto-mechanical assemblies allow to inject calibration light sources to the fiber link, for day calibrations as well as simultaneous RV referencing;
    \item Fiber selector allows switching between two science fibers set: 0.4” fiber in bright AO-assisted observations; 0.9” fiber for seeing-limited (or faint star AO);
    (Sect.~\ref{sec:guiding-system}).
\end{enumerate}
Those positions will be referenced back as [X] in the rest of the document.

\subsection{Interface to the telescope}

\begin{figure}
    \centering
    \begin{minipage}{0.53\textwidth}
    \centering
    \includegraphics[width=0.95\textwidth]{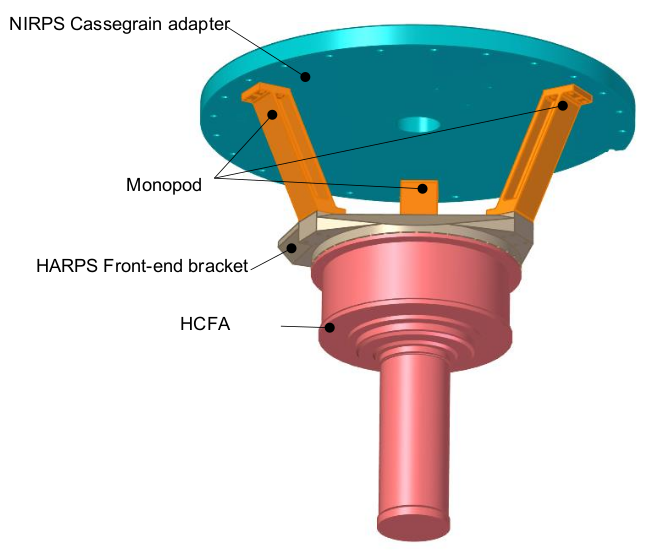}
    \caption{NFE mechanical structure. The NIRPS bench is mounted on the Cassegrain adaptor (seen in Fig.~\ref{fig:FE_pic}), while 3p6 TCCD/guiding functions have been relocated on the HCFA bracket.}
    \label{fig:structure}
    \end{minipage}
    \hfill
    \begin{minipage}{0.43\textwidth}
    \centering
    \includegraphics[width=0.95\textwidth]{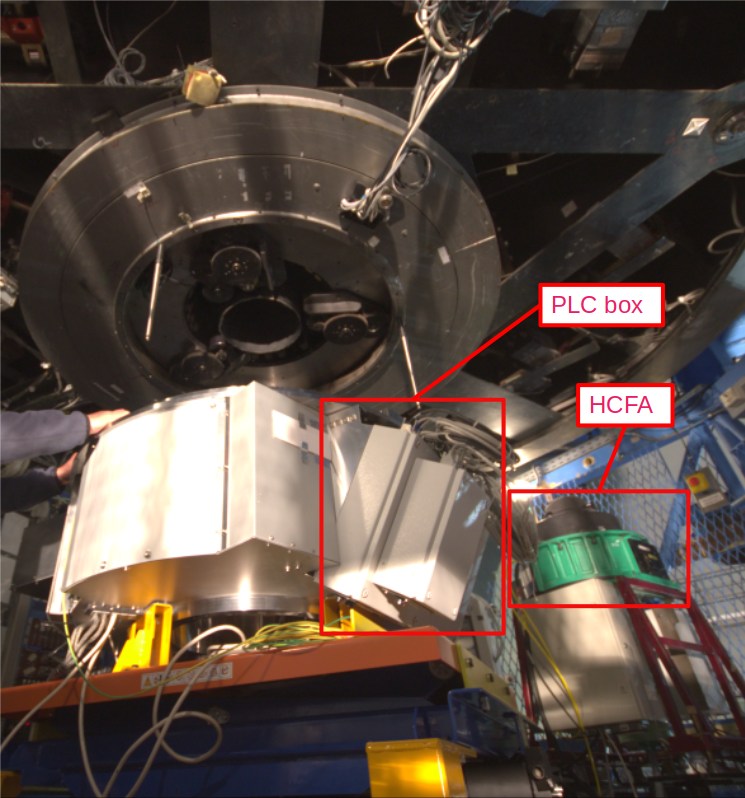}
    \caption{NFE with its protection panels during its installation to the 3p6. The HCFA is mounted on the NFE bottom bracket.}
    \label{fig:structure_pic}
    \end{minipage}
\end{figure}

The NFE is directly attached at the back of the 3p6 telescope rotator, between the telescope and the current HARPS Front-End (a.k.a. HCFA), in place of the old telescope adaptor. Fig.~\ref{fig:structure} shows the new tripod structure, entirely made of stainless-steel, like the optical mounts. NIRPS optics are positionned on the 'Cassegrain adapter' plate, while the HCFA is attached to the bottom bracket. The old adaptor contained in particular the 3p6 TCCD used for HARPS fast guiding, as well as a WFS for aligning the telescope after coating operations.
Those functions are now mechanically part of NFE (relocated on the bottom bracket), but they remain part of the HARPS/3p6 software. The TCCD has also been attached to a new XY motorized stage to preserve the off-axis guiding capability of the TCCD, with even higher reach. The check of the 3p6 alignement is also performed with the NFE WFS now.

The NIRPS/HARPS dichroic [1] is placed in the f/8.09 beam of the telescope. The dichroic has therefore been designed with a convex back surface (R=11m) to compensate for the defocus due to the 12mm glass thickness, to maintain the HCFA focal point position. The additional astigmatism generated by this optics is negligible compared to seeing for HARPS.

\subsection{Control system electronics}
All positioners are controlled via Beckhoff PLC system. A box containing the individual modules is directly attached to the NFE structure (Fig.~\ref{fig:structure_pic}), while the PLC computer is placed in the electronic cabinets, also in the cage, together with the RTC, DM control electronics, and guiding camera frame grabber.

\subsection{AO system}

The AO system consists in (see \cite{conod_2016a} for details):
\begin{itemize}
    \item A 14x14 Shack-Hartmann WFS, whose camera is a First Light Imaging OCAM2K, measuring WF between $\lambda =$ 700 and 950nm. With 15pixels/subap, the theoretical FoV is ~7", but is in practice limited to 4" by a field stop at FP2.
    \item An ALPAO DM241 High-Speed deformable mirror, providing up to 50$\mu m$ optical stroke and settling time as fast as 0.5ms;
    \item Hard-RTC is composed of rugged desktop components in a 3U 19" rack-mount case. All computations are performed on an Intel i9-10990X (8core) CPU, allowing computation of a new command vector in 300 $\mu s$. All operations considered, the RTC allows operation up to 1kHz.
    \item Soft-RTC runs on a Windows 10 OS, and through ALPAO A.C.E Matlab software. We give more information in Sec.~\ref{sec:soft-rtc};
\end{itemize}
In practice the AO system is operated at a frequency of 1kHz, with a measured loop delay of 3 frames. We choose a KL control basis, with up to 100 controlled modes for 136 actuators.

\subsection{Soft-RTC} \label{sec:soft-rtc}
The software part of the RTC runs on Matlab. The AO hardware (WFS \& DM) is interfaced thanks to the ALPAO ACE matlab libraries. The rest of the control software and interface to the ESO/VLT software  has been built on top.

Operation of the AO requires two Matlab instances to run in parallel. The two instances are interconnected via memory map files, which allow to exchange telemetry data, commands, etc, then allowing to control the AO and to forward system status to the instrument control software (ICS).
\begin{itemize}
    \item Server instance: The first Matlab instance is responsible of receiving commands coming from the ICS, check them, and forward them to the AO instance. It receives back telemetry data in open- and close-loop, which are then forwarded to the ICS for monitoring of AO status and behavior. 
    \item AO instance: this is the core real-time instance directly interfaced to the DM and WFS. It manages the command received from ICS and operates the system.
\end{itemize}
In order to avoid CPU resource conflict between the two instance during close-loop operations, we initially used the 'affinity' capability of W10 to assign CPU cores to each instance. It appeared to work well on the first version of our RTC (based on an Intel i9 gen 9) after disabling multi-threading, but not anymore on the latest commissioned version (Intel i9 gen 10), which appeared to work best with multi-threading and no particular affinity management.

\subsection{Guiding system}
\label{sec:guiding-system}
The NIR guiding camera looks at the field image reflected on the pierced mirrors of the fiber-head located at FP3 (Fig.~\ref{fig:FE_scheme}). Guiding is performed on the AO residual, located mostly outside the correction area of the DM.

The default plate scale is 30mas/pixel with a Field-of-view (FoV) of 18", allowing NCPA calibration at 1550nm with 3pix/fwhm sampling. A lens can be inserted to reduce the magnification by a factor of 3.3, for a plate scale of 100mas/pixel. At the end of the Front-End commissioning it appeared the low magnification is superfluous:
\begin{itemize}
    \item Guiding accuracy is improved by higher magnification, especially regarding the localization of the fiber hole;
    \item Tracking capabilities are still within specification on faint target;
    \item The 18" FoV is in practice enough to identify contaminants in TESS candidate fields;
    \item The telescope pointing accuracy is better than 5", so that the guide star can always be acquired in the FoV.
\end{itemize}
The camera is a First Light Imaging C-Red2. Conversely to most ESO systems, the guiding camera frames are directly collected by the instrument workstation (IWS) through the local network, with an intermediate PLEORA iPort CL-Ten frame grabber transferring the camera link frames to a dedicated line on the local network. The IWS then takes care of the frames bias and dark correction, and of measuring the star position from AO residual halo. The fiber head is subject to flexure drifts of up to 0.2" (< 10$\mu m$ in mechanical terms), so its position is also tracked. The newly computed fiber-star offsets are then sent to the AO system, that adjusts its reference slopes, as illustrated in Fig~\ref{fig:FE_scheme}. Every 10s the RTC estimates the average TT modes applied to the DM [4], and offload it to the TTM platform [5]. The platform gets into position in $\sim$0.2s, so that we did not implement a DM/TTM temporal decoupling.

Note that the NFE does not interact with the 3p6 telescope: the HCFA guiding will always be used for acquisition, even if HARPS is not operated. The tracking accuracy of the telescope is good enough, and TTM stroke and NFE FoV large enough to not have to implement another offload loop from the TTM to the telescope.

\section{Performance}

\begin{figure}[t!]
    \begin{minipage}{0.48\textwidth}
    \centering
    \includegraphics[width=0.95\textwidth]{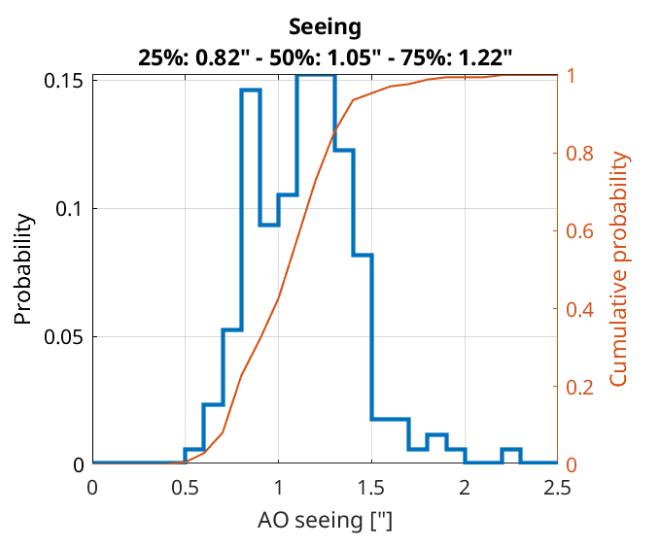}
    \end{minipage}
    \begin{minipage}{0.48\textwidth}\centering
    \includegraphics[width=0.95\textwidth]{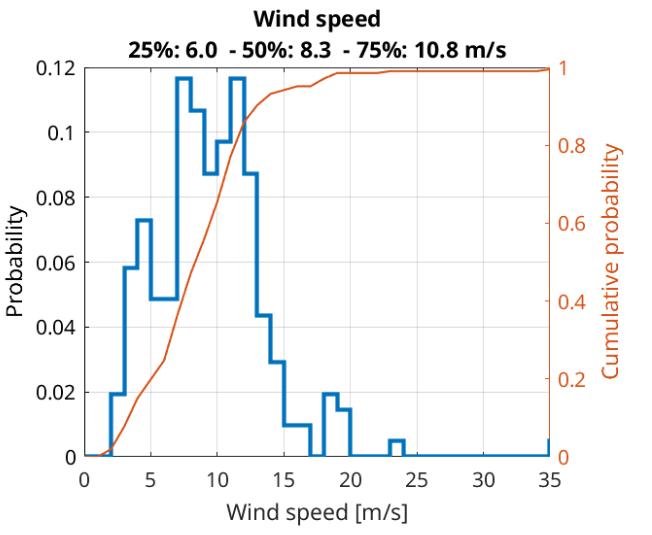}
    \end{minipage}
    \caption{Seeing conditions as estimated from the AO during last commissioning.}
    \label{fig:seeing}
    \vspace{0.5cm}
    \begin{minipage}{0.95\textwidth}\centering
    \includegraphics[width=0.8\textwidth]{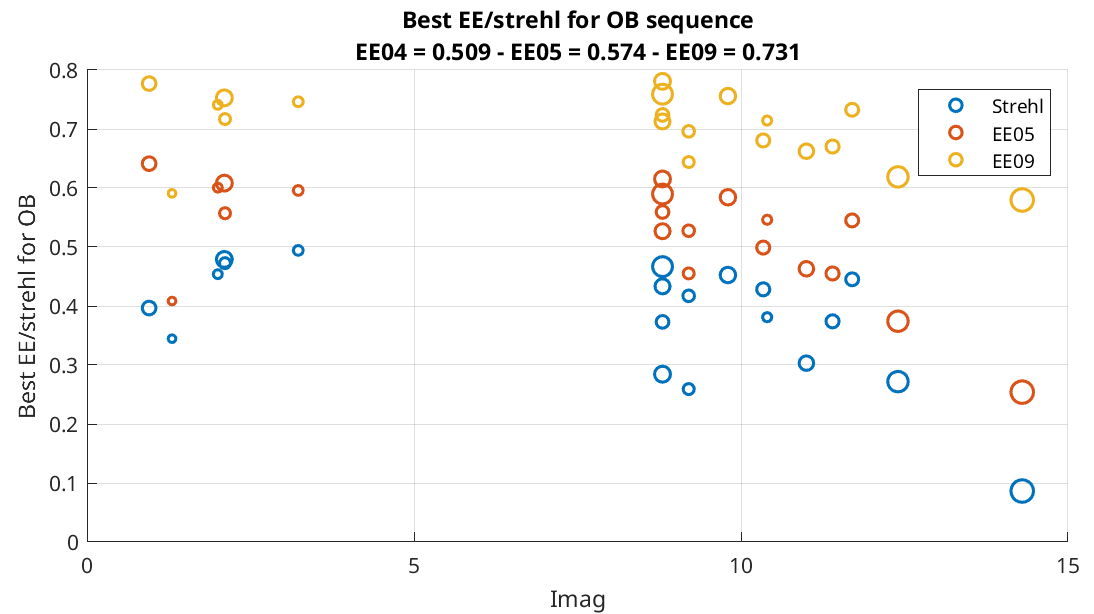}
    \caption{AO performance during last commissioning, with summarize on top mean performance on bright targets I$\ge$11. Bigger marker indicates better seeing (higher r0), between 0.7” and 2.2”.}
    \label{fig:ao_perf}
    \end{minipage}
\end{figure}

Fig.~\ref{fig:seeing} presents the seeing conditions estimated by the AO during last commissioning, with median seeing of 1.05", significantly worse than the design value of 0.85". The AO performance (Fig.~\ref{fig:ao_perf}) are estimated from 30s integrated guiding camera images. Note that the latter is un-filtered and therefore sees the whole band $\lambda$980-1700nm. The extracted values are Strehl($\lambda=1400nm$), and more important in our case, the Encircled Energy in 0.5" and 0.9" (noted EE05 and EE09). EE are reaching the expected values from simulations, reaching requirements up to I=11. Strehl reaches an average ~35\% on bright targets, while 55\% is expected from simulations in the observed median conditions. Tip-Tilt residuals measured by AO and those observed on guiding camera are perfectly in line, and in addition do not show any significant source of vibrations. 

The most intriguing observation is the lack of AO performance variations (Strehl in particular) for gain 0.05 to 0.4, or changing the number of controlled modes from Nkl=30 to 100, or the plateaued performance for all seeing and wind conditions. This suggests an external factor, unseen by the AO. The loss of coherence at all spatial frequencies is pretty obvious from the comparison of OTFs on-sky and on simulation in all observations.

The Strehl discrepancy between the measurements and simulations remains unexplained, but is consistent with an additional disturbance of the order of 130-170nm RMS, while the AO error budget is $\sim$100nm RMS in median seeing conditions.

\section{Lessons learned}

\begin{figure}[t!]
    \centering
    \includegraphics[width=0.8\textwidth]{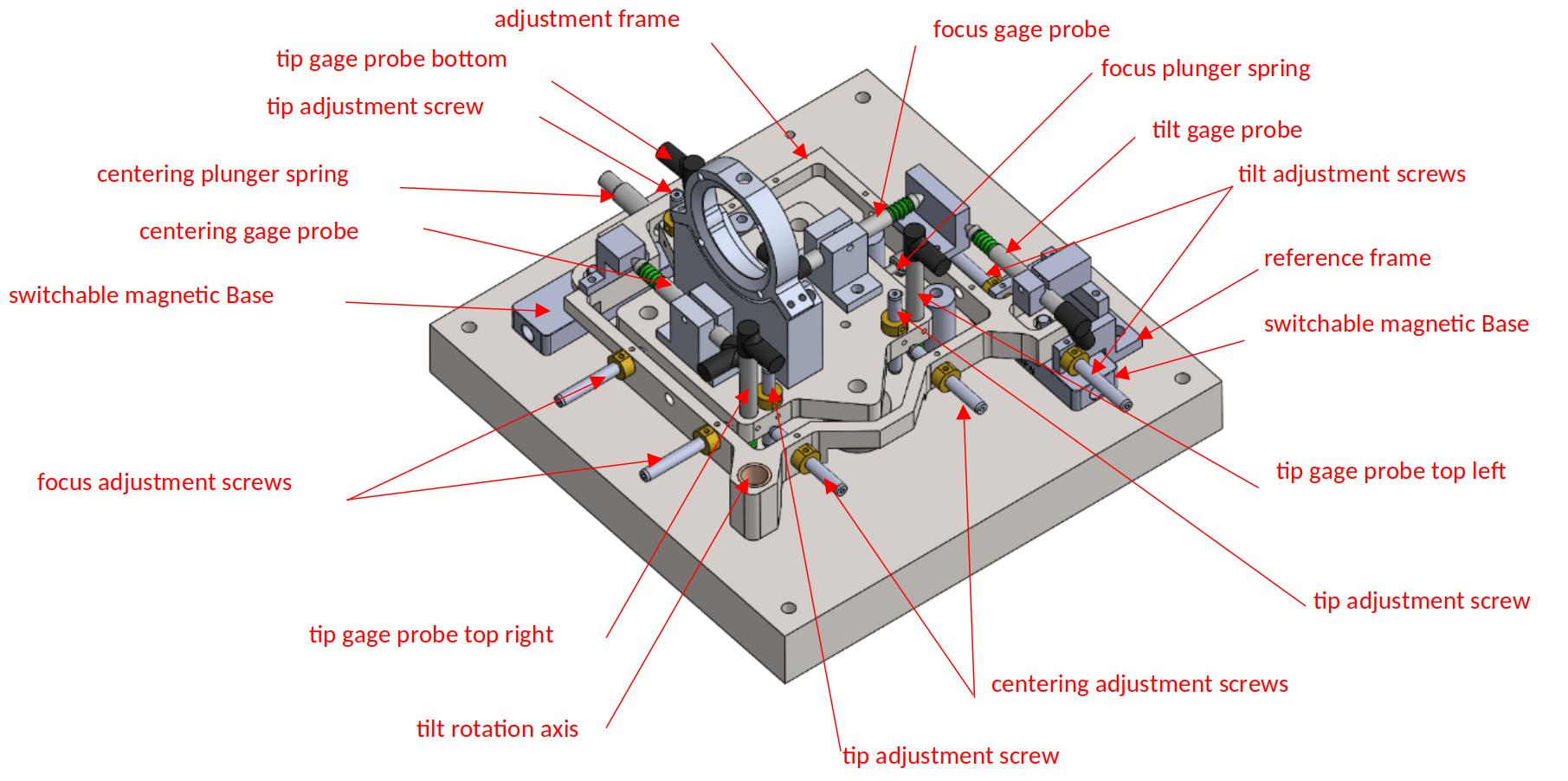}
    \caption{3D model of an optical mount with all positioning equipment.}
    \label{fig:alignment}
\end{figure}

\subsection{ADC wobbling}
\label{sec:adc}
The ADC assembly [3] of NIRPS follows the path of the ESPRESSO one, by having the double role of a classical ADC and of a focusing lens.
Unfortunately, this complex assembly was toleranced only regarding dispersion specifications, and resulted in too loose tolerances regarding the clocking error between the prisms of each pair. With an assembly error of 'only' $\sim 4^\circ$, the ADC generates important field ($\pm 6"$) \textbf{and} pupil wobbling ($\pm 0.5$ subaperture on WFS). We could verify that each ADC was contributing to about 50\% of the effect, each giving rise to its own independent circular wobbling pattern. 
ADC wobbling forced to motorize the M3 fold mirror close from the focus, in order to stabilize the pupil. M3 being 50mm in-front of focal point, the pupil stabilization also generates some significant level of field motion, of the order of 1" per subaperture.

A TTM [5] correction law can be calibrated in the NFE itself, mostly to improve the efficiency of the acquisition sequence. In operation, the guiding system and DM offload to the TTM take care of the wobbling tracking. Extra-care has been taken to slow down the ADC correction when passing near zenith, where ADCs can rotate and wobble faster than the offload loop can manage. The ADC correction error is negligible in this area.

The PTTM [2] correction law to keep the pupil centered cannot be calibrated in the NFE, the optics being in front of FP1. It was necessary to calibrate it in the lab during AITs. We did so by placing a laser beacon 20m away from the NFE FP1, i.e. at the position of the 3p6 pupil, and imaging the pupil/beacon and DM on a lab camera. During AIV in La Silla, a constant offset value was added to the laws to center the pupil on the WFS. The laws calibrated in early 2019 were finally validated in October 2021.

More details on the ADC design and integration can be found in \cite{cabral_2022a}.

\subsection{DM241 creep and heating}

Like in other experiments like NAOMI@VLTI, we could observe strong creeping and heating on our DM241. Placing a temperature sensor on the back of the DM, we could observe significant heating building up over few minutes, and much slower cool down ($\ge$ 15min). We implemented an open-loop sine-modulation of DM focus at 250Hz to monitor the evolution of the DM characteristics, in particular a change in the response of actuators, assuming all actuators behave the same. Thorough characterization of this effect could not be performed during AITs. We nevertheless could not notice performance degradation during 30min comm observations, either from the DM monitor itself or from the guiding camera Strehl and EE (which seem limited by an external factor).

Due to a misuse of the NFE, the DM was left in a heat-generating configuration, and we lost an actuator, clearly visible as a noisy column in the IM. This actuator is nevertheless placed on the edge of the secondary obstruction, and thanks to the ALPAO voice-coil technology, it is freely dragged by its neighbours. We could not notice an impact on performance.

\subsection{Alignment system}
To simplify the alignment, and following ESPRESSO experience, we performed the optical alignment based only on manufacturing and positioning tolerances. After reception of optics and manufacturer reports, the optical design and positions were recomputed for the final alignment.
A common interface plate was designed for all opto-mechanics. A dedicated positioning jig was developed, allowing 5 DoFs micro-adjustments (linear stroke of $\pm$ 2mm), monitored with Linear Variable Differential Transformer gauge probes (Fig.~\ref{fig:alignment}). After measuring the plate position with a portable CMM, corrections are computed for each DoFs. The operation is repeated until the residual errors are acceptable, and then the position is locked with shims and eccentric end-stops. The system has demonstrated positioning accuracy $\leq 10\: \mu m$ and re-positioning accuracy $\sim 20\: \mu m$, up to 5 times better than required.

We comment here about some of the difficulties we encountered:
\begin{itemize}
\item A too high confidence on the alignment of our doublets after transport undermined the performance of this strategy. It appeared that the FP2 \& FP3 focusing doublets suffered from a lateral misalignment of $\sim 50 \mu m$, leading to significant coma aberration when positioning using sole mechanical interfaces. The aberration was corrected by applying a $\sim 1^\circ$ angle to the doublet assembly, but it strongly displaces the focal points positions. An XY position correction of the doublets was then performed to recover the focal points, but that motion exceeded the foreseen capability of one of the platforms, requiring re-machining. This correction procedure was first performed visually, before measuring the optics with our interferometer and updating the ideal position in Zemax and the positioning system software. The absence of a few reference targets and of convenient interfaces for test equipment increased the complexity of this correction procedure.
\item The alignment of the guiding camera 'zoom' optics (Sect.~\ref{sec:guiding-system}) had to be optimized by hand. The camera position was optimized for the fixed optics as planned. Due to small error in the effective focal length of the 'insertable' doublet, its focal point was different: the positioning of the insertable optics was then performed visually, looking at the image quality on the camera.
\item We initially used a dummy fiber head to reference FP3. This dummy was only designed for fiber tests, and was therefore not mechanically referenced. This led to some iterations before we find the proper FP3 position, leading to proper focus on the guiding camera.
\end{itemize}

\bibliography{references} 

\begin{thebibliography}{1}

\bibitem{conod_2016a}
Conod, U. et~al., ``Adaptive optics for high resolution spectroscopy: a direct
  application with the future {NIRPS} spectrograph,''  990941 (July 2016).

\bibitem{cabral_2022a}
Cabral, A. et~al., ``Nirps – the near infra red planet searcher: design,
  integration and tests of the atmospheric dispersion compensator,''  {\bf This
  proceeding},  990941 (Aug. 2022).

\end{thebibliography}
\bibliographystyle{spiebib} 

\end{document}